\documentclass[prl,twocolumn,showpacs,superscriptaddress]{revtex4}
\usepackage[OT1]{fontenc}
\usepackage[latin1]{inputenc}
\usepackage{color}
\usepackage{graphicx}
\usepackage{amssymb,amsmath}
\usepackage{upgreek}
\begin{document}

\title{Doppler cooling to the Quantum limit}
\author{M. Chalony}
\affiliation{Institut Non Lin\'eaire de Nice, Universit\'e de Nice Sophia-Antipolis, CNRS, 06560 Valbonne, France}
\author{A. Kastberg}
\affiliation{Laboratoire de Physique de la Mati\'ere Condens\'ee, Universit\'e de Nice Sophia-Antipolis, UMR 6622, Parc Valrose, 06108 Nice Cedex 2, France}
\author{B. Klappauf}
\affiliation{Department of Physics and Astronomy, UBC 6224 Vancouver, Canada}
\author{D. Wilkowski}
\affiliation{Institut Non Lin\'eaire de Nice, Universit\'e de Nice Sophia-Antipolis, CNRS, 06560 Valbonne, France}
\affiliation{Centre for Quantum Technologies, National University of Singapore, 117543 Singapore, Singapore}
\affiliation{PAP, School of Physical and Mathematical Sciences, Nanyang Technological University, Singapore 637371, Singapore}
\date{\today{}}

\begin{abstract}
Doppler cooling on a narrow transition is limited by the noise of single scattering events. It shows novel features, which are in sharp contrast with cooling on a broad transition, such as a non-Gaussian momentum distribution, and divergence of its mean square value close to the resonance. We have observed those features using 1D cooling on an intercombination transition in strontium,  and compared the measurements with theoretical predictions and Monte Carlo simulations. We also find that for a very narrow transition, cooling can be improved using a dipole trap, where the clock shift is canceled.
\end{abstract}


\maketitle

Laser cooling of atoms is a technique widely used, often as a first cooling stage on the road to quantum degeneracy. In the framework of Doppler cooling, a moving atom is cooled because of the difference in absorption probabilities, induced by the Doppler effect, between quasi-resonant red-detuned counterpropagating laser beams \cite{hansch1975cooling,windeland1975cooling}. At each scattering event the momentum is, in average, modified by a recoil unit $\hbar k$, with $k$ the wave number. If the steady state root mean square (rms) of the momentum distribution,
\begin{equation}
\sigma_p\gg\hbar k \, ,
\label{eq_semi_class}
\end{equation}
is much larger than the recoil momentum. Doppler cooling can be expanded into a semi-classical theory, where the momentum evolution follows a damped Brownian motion with a pure friction force. In the low intensity regime the distribution is thermal-like with,
\begin{equation}
\sigma_p^2=mk_bT=\frac{7m\hbar}{80|\delta |}(4\delta^2+\Gamma^2) \, ,
\label{eq_semi_Doppler}
\end{equation}
for a standing wave in 1D. $\delta$ is the laser detuning, $\Gamma$ the linewidth of the transition and $m$ the atomic mass. From the relation (\ref{eq_semi_Doppler}), the inequality (\ref{eq_semi_class}) can be reformulated as
\begin{equation}
\omega_\mathrm{r}=\frac{\hbar k^2}{2m}\ll\Gamma \, ,
\label{eq_broad_trans}
\end{equation}
where $\omega_\mathrm{r}$ is the recoil frequency. Thus the semiclassical Doppler theory, discussed above, is valid only for atoms having a broad transition, i.e. fulfilling the inequality (\ref{eq_broad_trans}).

In the 1980s, with the progress of laser cooling and trapping techniques, in parallel with precise measurements of the momentum distribution, Doppler theory was found to be too crude to explain the steady state regime of atoms with complex internal structure. The ground state Zeeman manifold turned out to play a crucial role leading to "sub-Doppler cooling" \cite{lett1988observation,shevy1989proceeding,dalibard1989proceeding,dalibard1989laser,ungar1989optical}. However, Doppler theory remains valid for a two-level atom and for $J=0\rightarrow J=1$ systems. Several attempts were made to quantitatively test the theory of a broad transition, using spin polarized alkaline atoms \cite{weiss1989optical}, or more recently with spinless, ground state, bosonic two-electrons atoms \cite{PhysRevA.66.011401,chaneliere2005extra,cristiani2010fast,mcferran2010sub}. However, most of those measurements show higher temperatures than predicted by the theory. Explanations suggested include spurious heating effects coming from spatial fluctuations of the laser intensity \cite{chaneliere2005extra}, and excited state coherences \cite{PhysRevA.77.015405}.

Two-electron atoms remain an ideal testing ground for Doppler theory. In addition to broad transitions, narrow intercombination transitions also exist, where the inequality (\ref{eq_broad_trans}) no longer holds. Those later cases lead to new features of the steady state regime, theoretically studied by Castin and co-authors using a full quantum approach of the Doppler theory \cite{castin1989ldc}. In particular, close to the resonance, but still on the red side of the transition, the momentum distribution has a non-Gaussian shape characterized by long tails, leading to a divergence of the rms momentum. Hence the minimum rms momentum -lower bounded by the recoil momentum- is red shifted in frequency, with respect to the $\delta=-\Gamma/2$ value for broad transitions. The use of two-electron atoms in laser cooling with narrow intercombination lines is growing, and several groups already reported laser cooling on narrow transitions with temperature at or close to the recoil limit \cite{katori1999magneto,curtis2001quenched,loftus2004narrow,poli2005cooling,chaneliere2008three,PhysRevLett.100.113002,ido2000optical,grain2007feasibility}. Most of those works were carried on in a magneto-optical trap (MOT) where the cooling dynamic and steady state regime are modified by the presence of the magnetic field gradient \cite{loftus2004narrow,chaneliere2008three}. Moreover, the gravity plays an important role displacing the position of the cloud and thus changing its magnetic environment. In this letter, we present some experiments done using a dipole trap where the clock shift is canceled and we discuss the first quantitative test of this important theory of quantum Doppler cooling.

\begin{figure}
\begin{center}
\includegraphics[scale=0.45]{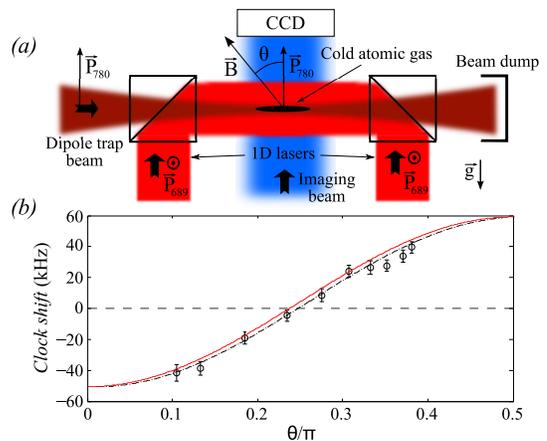}
\caption{\emph{(a)}: Outline of the 1D experimental set-up. The cold atomic gas, at the center of the picture, is held by the focused far-off resonance laser beam at $780$~nm. The 1D cooling lasers at $689$~nm are superimposed on the $780$~nm beam. Along a perpendicular axis, an imaging system on the $461$nm broad transition records the spatial distribution of the cloud. A $0.3$~G magnetic bias field is adjusted in angle $\theta$ with respect to the polarization of the $780$~nm beam in order to cancel the clock shift on the 1D laser system. \emph{(b)}: Variation of the clock shift as a function of $\theta$. The circles are experimental data points, whereas the full red curve is the predicted behavior. The dashed line corresponds to the red curve shifted in $\theta$ to show that the small disagreement between the experimental point and the predicted value is most likely due to a systematic error in the angle calibration.}
\label{set-up_Clock-shift}
\end{center}
\end{figure}

As described in \cite{chaneliere2008three}, our $^{88}$Sr cold atomic sample is produced as follow; after a loading and a precooling stage on the broad $^1\!S_0\rightarrow\,^1\!P_1$ dipole-allowed transition ($\Gamma/2\pi=32$~MHz, $\omega_\mathrm{r}/2\pi=10.6$~kHz), the atoms are transferred to a MOT operating on the narrow $^1\!S_0\rightarrow\,^3\!P_1$ intercombination transition at $689$ nm ($\Gamma/2\pi=7.5$~kHz, $\omega_\mathrm{r}/2\pi=4.7$~kHz). This final cooling stage lasts for $130$~ms and leads to a cold gas containing about $2\times 10^7$ atoms at a temperature of $T=2\,\mu$K. A few tens of milliseconds before switching off the MOT, a far-off resonant dipole trap, centered on the atomic gas, is turned on. This dipole trap consists of a single focused laser beam at $780$ nm. The laser power is $120$~mW for a beam waist of $17\,\mu$m, corresponding to a potential depth of $T_0\simeq 20\,\mu$K. The radial and longitudinal trap frequencies are respectively $\omega_\perp=670$~Hz and $\omega_\parallel=8$~Hz. Because of the weak overlap between the dipole trap and the initial cold cloud, at best $1\%$ of the atoms are transferred into the dipole trap. $50$~ms after the MOT stage, a counter-propagating pair of beams, aligned with the long axis of the dipole trap and red-detuned with respect to the $^1\!S_0\rightarrow\,^3\!P_1$ transition is turned on for $450$~ms (Fig. \ref{set-up_Clock-shift}). We carefully balance the intensity of the cooling beam minimizing the displacement of the gas center of mass in the dipole trap. The cloud's spatial distribution is recorded by absorption imaging on the broad line at $461$~nm.

%%%%%%%%%%%%%%%%%%%%%%%%%%%%%%%%%%%%
%%%%%%%%%%%%%%%%%%%%%%%%%%%%%%%%%%%%

A $0.3$~G magnetic bias field (\emph{\textbf{B}}) is applied during the 1D cooling experiment, for two important reasons. First, the Zeeman degeneracy of the excited state is lifted so that the lasers interact only with a two-level system made out of the $m=0\rightarrow m=0$ transition, which is unsensitive to residual magnetic field fluctuations. Second, the orientation of \emph{\textbf{B}}, with respect to $\textbf{P}_{780}$, the linear polarization of the dipole trap beam, is tuned to cancel the clock (or transition) shift induced by the dipole trap on the transition of interest \cite{PhysRevLett.91.053001}. The variation of the clock shift with respect to the angle $\theta$ between $\textbf{\emph{P}}_{780}$ and \emph{\textbf{B}} is given in Fig. \ref{set-up_Clock-shift}b, and is in good agreement with the predicted value. The relative accuracy of the clock shift cancellation, for the whole trapping potential, is about $4$~kHz, \emph{i.e.}, below the bare transition linewidth $\Gamma$. At this precision, one can ignore the small spatial dependency of the laser detuning induced by the dipole trap over the confined atomic gas. The position of the resonance has been measured with a precision of $\pm\Gamma/2$ using absorption spectroscopy on the cold cloud. For technical reasons, the linear polarization $\textbf{\emph{P}}_{689}$ of the cooling lasers is not aligned with the bias B-field. Thus the effective coupling intensity is reduced by a factor $\cos^2{\alpha}\simeq 1/15$, where $\alpha$ is the angle between $\textbf{\emph{P}}_{689}$ and $\textbf{\emph{B}}$. Taking into account this reduction factor, the effective intensity $I_f$ is in the range $0.03-0.1 I_s$, where $I_s=3\,\mu\textrm{W/cm}^2$ is saturation intensity. At those intensities, the damping times of the momentum are long, in the range $0.02-0.2$~s. As a consequence, the steady state regime is usually not reached in a free falling experiment, which validates the use of the trapping potential.

%%%%%%%%%%%%%%%%%%%%%%%%%%%%%%%%%%%
\begin{figure}
\begin{center}
\includegraphics[scale=0.5]{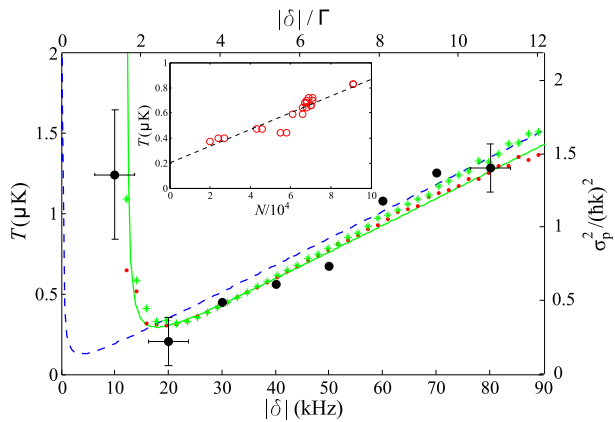}
\caption{Mean square momentum in temperature and recoil units as function of cooling laser detuning. The experimental data (black, full circles), extrapolated for $N\rightarrow0$ (noninteracting gas), are compared to the full quantum approach developed by Castin \emph{et al.}\cite{castin1989ldc} (green, solid line). The blue, dashed curve shows the prediction for broad transition Doppler theory. The red points (green stars) correspond to the MC simulation performed in (without) the dipole trap. \emph{Inset}: Measured mean square momentum as function of the number of atoms in the trap with $\delta=-20$~kHz and $I_f=0.06I_s$ (red circles) with linear fit (dashed line).
}
\label{temperature}
\end{center}
\end{figure}
%%%%%%%%%%%%%%%%%%%%%%%%%%%%%%%%%%%%

In Fig. \ref{temperature} we show the experimental mean square momentum (black full circles) in temperature and recoil units. The comparison with the analytical prediction of the full quantum approach \cite{castin1989ldc} (green curve) shows an excellent agreement. Signatures of a quantum nature of  Doppler cooling are found, \emph{e.g.}, the mean square momentum has a minimum below the recoil, and it shows a divergence close to the resonance, but at a detuning clearly larger negative than $\delta=-\Gamma/2$ predicted by the broad transition Doppler theory (see [Eq. (\ref{eq_semi_Doppler})] and the corresponding plot: blue, dashed curve in Fig. \ref{temperature}). Spurious heating effects reported for broad transitions \cite{chaneliere2005extra,PhysRevA.77.015405} are not observed, most likely because we are using a two-level system with a long damping time.

Several steps are necessary in order to validate the experiment and theory comparison of the preceding paragraph. First, the value of the mean square momentum can not be deduced from the spatial expansion of the cloud in the trap at the steady state regime, because of the presence of light induced collective effects. For a standard 3D magneto-optical trap, the cloud would inflate due to the repulsive multiple scattering force \cite{walker1990collective}. In contrast, we observe a compression of the cloud because of the dominance of the attractive shadow force in the 1D configuration \cite{dalibard1988laser,autoG}. The mean square momentum is then deduced by measuring the evolution of the cloud over half of the trap period, after switching off the 1D cooling laser. For that purpose, we use the relationship
$\sigma_z^2(t)= \sigma_z^2(0)\cos^2{(\omega t)}+\left(\sigma_p(0)/m\omega \right)^2\sin^2{(\omega t)} \, ,$
linking the rms value of the cloud size in the 1D harmonic trap, $\sigma_z(t)$, to the initial values (at the switching off of the lasers) in real space and momentum space.

Moreover, the mean square momentum depend on the number of atoms, showing that collective effects also induce extra heating. Regardless of the exact origin of this extra heating, one example of this dependency is shown in the inset of Fig. \ref{temperature}. We extrapolate the mean square momentum value to the noninteracting limit (vanishing number of atoms) using a linear fit. The data points for the mean square momentum in Fig. \ref{temperature} are deduced in this way.

Finally, even if the trap is loose along the cooling axis ($\omega_\parallel\ll\omega_\mathrm{r}$), it is not clear that it does not affect the cooling process. Later on we will show that the trapping indeed has a major impact when the transition is very narrow, $\omega_\mathrm{r}\gg\Gamma$. To explore the influence of the trap, we use a Monte Carlo (MC) simulation comparing cooling with and without the trap. The MC simulation is based on a rate equation describing the scattering events where the external degrees of freedom are treated classically. This approach, neglecting the quantum nature of the external degrees of freedom, is known to be consistent with the full quantum approach in free space \cite{castin1989ldc}. This point is also confirmed here, where the results of the MC simulation in free space are plotted in Fig. \ref{temperature} (green stars). In the trap, the dynamic is more subtle because of the presence of the trapping force. It turns out, however, that for the strontium intercombination transition with $\omega_\mathrm{r}\simeq0.6\Gamma$ the trap does not significantly modify the mean square momentum in the steady state (red dots in Fig. \ref{temperature}). We will see later that the condition $\omega_\mathrm{r}\gg\Gamma$ leads to different conclusions.

%%%%%%%%%%%%%%%%%%%%%%%%%%%%%%%%%%%
\begin{figure}
\begin{center}
\includegraphics[scale=0.62]{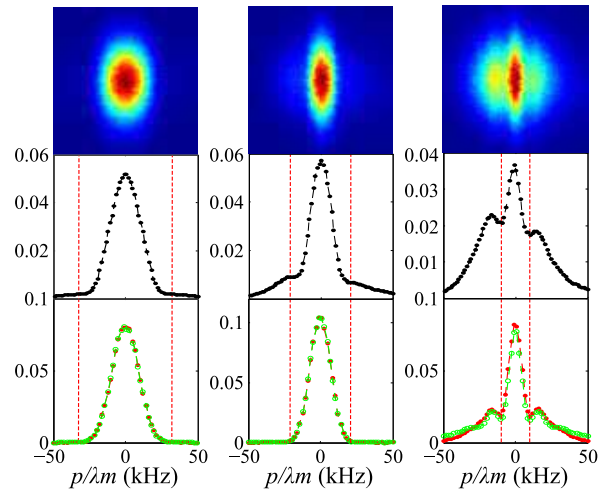}
\caption{Upper panels: Raw data, false color image of the atomic cloud after 1D cooling in free space, and $50$~ms time of flight. The cooling laser are along the horizontal axis. From left to right the laser detunings are respectively $\delta=-33$~kHz, $\delta=-21$~kHz and $\delta=-10$~kHz, and the laser intensity is around $0.5I_s$. The small horizontal asymmetry is likely due to imperfect balance of the cooling beams intensity, which were not precisely adjusted like for the cloud in the Dipole trap case. Middle panels: Normalized spatial distribution along the cooling axis extracted from the upper images. Lower panels: Normalized momentum distribution extracted from the MC simulation. The laser detuning is the same as in the experiment, but the simulation is performed at the low intensity limit. These plots show a qualitative agreement with the experiment at higher intensity without the added trap (green open circles), as well as in the trap (red dots). The resonance lines correspond to the vertical, dashed lines.}
\label{distribution}
\end{center}
\end{figure}
%%%%%%%%%%%%%%%%%%%%%%%%%%%%%%%%%%%%

Other signatures of the quantum nature of Doppler cooling can be found in the shape of the momentum distribution. In the broad transition semiclassical picture, the momentum distribution is essentially Gaussian since it remains very well confined far from the two $\pm\delta m/k$ resonance lines. With a narrow transition, a single scattering event might be enough to bring the atom out of resonance. As a consequence, the momentum distribution is not Gaussian anymore and shows out-of-resonance long tails and dips at the resonance lines \cite{castin1989ldc,yoo1991wigner}. A precise measurement of the momentum distribution has been done for the case of free space 1D cooling on clouds with large number of atoms, to improve the signal to noise ratio. The laser intensity was increased to $0.5I_s$ in order to reach the steady state regime during the laser interaction time. This is a likely explanation to why there is only a qualitative agreement found between the experiment and the MC simulations done for the low intensity limit (see Fig. \ref{distribution}). The experimental momentum distributions (Fig. \ref{distribution}, middle panels) might be decomposed into two domains; between and beyond the two $\pm\delta m/k$ resonance lines (red, dashed, vertical lines). The distribution on the inside is in very good agreement with the MC simulation since it depends only on the laser detuning. For the contribution outside the resonance lines, the dependence is more pronounced for the experimental cases, revealing that extra heating is at play.

%%%%%%%%%%%%%%%%%%%%%%%%%%%%%%%%%%%
\begin{figure}
\begin{center}
\includegraphics[scale=0.62]{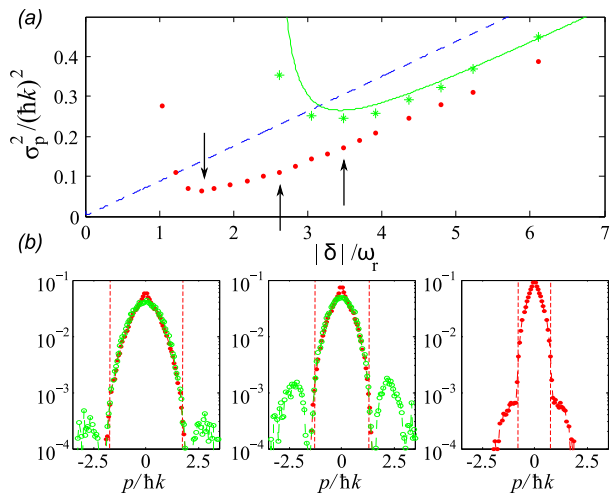}
\caption{MC simulation on the calcium intercombination transition. \emph{(a)}: Mean square momentum in recoil units, as function of the cooling laser detuning. The red points (green crosses) correspond to the MC simulation performed in (without) the dipole trap. The green, full line and the blue, dashed line correspond, respectively, to the analytical full quantum expression \cite{castin1989ldc}, and broad transition limit ([Eq.~(\ref{eq_semi_Doppler})]). \emph{(b)}: Normalized momentum distribution without trap (green open circles), and in the trap (red dots). From left to right the laser detunings are respectively $\delta=-3.5\omega_\mathrm{r}$, $\delta=-2.6\omega_\mathrm{r}$ and $\delta=-1.6\omega_\mathrm{r}$. Black arrows indicate those points in \emph{(a)}. The resonance conditions correspond to the dashed vertical lines.}
\label{distribution_Ca}
\end{center}
\end{figure}
%%%%%%%%%%%%%%%%%%%%%%%%%%%%%%%%%%%%

So far, we have discussed Doppler Cooling on the narrow strontium intercombination transition with $\Gamma\simeq\omega_\mathrm{r}\simeq k\sigma_p/m$. We have shown that the trap has no major impact on the steady state regime. We will now consider the case of a very narrow transition, \emph{i.e.} $\Gamma\ll\omega_\mathrm{r}\simeq k\sigma_p/m$, where the laser excitation is well localized in the momentum space. In the trap, the laser excitation frequency is broaden by the oscillation of the atom and occurs above an energy threshold corresponding to $\hbar|\delta|$. This configuration has strong similarity with the broadband cooling proposal discussed in Ref. \cite{wallis1989blc}. It is expected to be more efficient than single frequency cooling, since an atom outside the resonance lines can still be cooled and brought back to the central region. As a consequence the long tails are reduced in  the trap with respect to the free space case. This effect is in fact evident also for the strontium transition (see lower-right panel on Fig. \ref{distribution}). However, it will be more pronounced with a very narrow transition. As an example, MC simulations were performed for the calcium intercombination transition ($\lambda=657$~nm, $\Gamma/2\pi=400$~Hz and $\omega_\mathrm{r}/2\pi=11$~kHz), using the same dipole trap parameters as previously described (see Fig. \ref{distribution_Ca}). We  observed that the momentum distribution is more confined in the central region, leading to a reduction of its mean square value by more than a factor of 3, with a minimum closer to the resonance line. More systematic studies are left for future investigation.

To conclude, we have found an excellent agreement between cooling on the narrow intercombination transition in strontium, and the quantum theory of Doppler cooling developed in \cite{castin1989ldc}. As a major feature of cooling on narrow transitions, the momentum distribution can be decomposed into a cold part at lower momenta than one corresponding to the laser resonance line, and a hotter part at higher momenta. This latter component can be strongly reduced by using a very narrow transition in a dipole trap, where the clock shift is canceled.

%\bibliography{Narrow_Cooling}
%\begin{thebibliography}{18}
%\expandafter\ifx\csname natexlab\endcsname\relax\def\natexlab#1{#1}\fi
%\expandafter\ifx\csname bibnamefont\endcsname\relax
%  \def\bibnamefont#1{#1}\fi
%\expandafter\ifx\csname bibfnamefont\endcsname\relax
%  \def\bibfnamefont#1{#1}\fi
%\expandafter\ifx\csname citenamefont\endcsname\relax
%  \def\citenamefont#1{#1}\fi
%\expandafter\ifx\csname url\endcsname\relax
%  \def\url#1{\texttt{#1}}\fi
%\expandafter\ifx\csname urlprefix\endcsname\relax\def\urlprefix{URL }\fi
%\providecommand{\bibinfo}[2]{#2}
%\providecommand{\eprint}[2][]{\url{#2}}

\end{document}